\documentclass[aps,superscriptaddress,preprintnumbers,showpacs,nofootinbib,floatfix]{revtex4}
\usepackage{graphicx,epsfig,wrapfig,amssymb}
\newcommand{\be}{\begin{equation}}
\newcommand{\ee}{\end{equation}}
\newcommand{\ba}{\begin{eqnarray}}
\newcommand{\ea}{\end{eqnarray}}
\newcommand{\la}{\langle}
\newcommand{\ra}{\rangle}
\newcommand{\di}{ {\rm d} }
\newcommand{\limNonRel}{{{\lim_{\mbox{\tiny non-rel}\,}}}}
\begin{document}
%===================  TITLE, AUTHORS, AFFILIATIONS ===================
%\newcommand*{\xxx}{bla, bla}\affiliation{\xxx}
\newcommand*{\Dubna}{
	Bogoliubov Laboratory of Theoretical Physics, JINR, 141980 Dubna, 
	Russia}\affiliation{\Dubna}
\newcommand*{\UConn}{
        Department of Physics, 
        University of Connecticut, Storrs, CT 06269, USA}\affiliation{\UConn}
\newcommand*{\Praha}{
	Institute of Physics, Academy of Sciences of the Czech Republic, 
	Na Slovance 2, CZ-182 21 Prague 8}\affiliation{\Praha}
\title{Transverse momentum dependent distribution functions \\
      in a covariant parton model approach with quark orbital motion}
\author{A.~V.~Efremov}\affiliation{\Dubna}
\author{P.~Schweitzer}\affiliation{\UConn}
\author{O.~V.~Teryaev}\affiliation{\Dubna}
\author{P.~Zavada}\affiliation{\Praha}

\date{March, 2009}%\date{\today}
%===================  PREPRINT NUMBER, JOURNAL =======================
%\preprint{\tt preprint: }
%===================  ABSTRACT =======================================
\begin{abstract}
\noindent
Transverse parton momentum dependent distribution functions (TMDs) 
of the nucleon are studied in a covariant model, which describes the 
intrinsic motion of partons in terms of a covariant momentum distribution. 
The consistency of the approach is demonstrated, 
and {\sl model relations} among TMDs are studied.
As a byproduct it is shown how the approach allows to formulate 
the non-relativistic~limit.
\end{abstract}
 % Hadron-induced inclusive production with identified leptons,
                % photons, or other nonhadronic particles (energy > 10 GeV)
%\keywords{Semi-inclusive deep inelastic scattering,
% transverse momentum dependent distribution functions}
\maketitle
%===================  SECTION 1: INTRODUCTION ========================
\section{Introduction}
\label{Sec-1:introduction}

Studies of hard scattering processes such as inclusive deep-inelastic lepton 
nucleon scattering (DIS) have given rise to a good understanding of parton 
distribution functions, which tell us how the parton momenta parallel to the 
nucleon momentum are distributed. One way to gain insights into the partonic 
quark-gluon substructure of the nucleon 
beyond this one-dimensional picture is to consider transverse momentum 
dependent ('unintegrated') parton distributions (TMDs) \cite{Collins:2003fm}.
These objects can be accessed by observing transverse momenta of, e.g., 
hadrons produced in semi-inclusive DIS (SIDIS) or dileptons produced 
in the Drell-Yan process 
\cite{Cahn:1978se,Sivers:1989cc,Efremov:1992pe,Kotzinian:1994dv,Mulders:1995dh,Boer:1997nt,Boer:1999mm,Brodsky:2002cx,Collins:2002kn}
thanks to factorization
\cite{Efremov:1980kz,Collins:1981uk,Ji:2004wu,Collins:2004nx}.
Much theoretical progress was made 
\cite{Bacchetta:1999kz,Belitsky:2002sm,Metz:2002iz,Pobylitsa:2003ty,Brodsky:2006hj,Avakian:2007xa,Burkardt:2007rv,Miller:2007ae,Goeke:2005hb,Bacchetta:2006tn,Avakian:2007mv,Metz:2008ib},
and first data
\cite{Arneodo:1986cf,Airapetian:1999tv,Airapetian:2001eg,Airapetian:2002mf,Avakian:2003pk,Airapetian:2004tw,Alexakhin:2005iw,Diefenthaler:2005gx,Gregor:2005qv,Ageev:2006da,Avakian:2005ps,Airapetian:2005jc,Airapetian:2006rx,Abe:2005zx,Ogawa:2006bm,Martin:2007au,Diefenthaler:2007rj,Kotzinian:2007uv,Seidl:2008xc}
give rise to phenomenological insights 
\cite{DeSanctis:2000fh,Anselmino:2000mb,Efremov:2001cz,Efremov:2001ia,Ma:2002ns,Efremov:2003eq,D'Alesio:2004up,Anselmino:2005nn,Efremov:2004tp,Collins:2005ie,Collins:2005rq,Vogelsang:2005cs,Efremov:2006qm,Anselmino:2007fs,Arnold:2008ap,Anselmino:2008sg,Kotzinian:2006dw}.

Nevertheless, presently model studies 
\cite{Jakob:1997wg,Avakian:2008dz,Pasquini:2008ax,Efremov:2008mp,She:2009jq,Bacchetta:2008af,Meissner:2007rx,Bacchetta:2002tk,Efremov:2002qh,Gamberg:2003ey,Yuan:2003wk,Lu:2004au,Burkardt:2007xm,Gamberg:2007wm,Courtoy:2008vi,Boffi:2009sh}
play an important role.
It is worth to recall that important insights concerning the very existence
\cite{Brodsky:2002cx} or universality \cite{Metz:2002iz} of effects 
were made on the basis of model studies, see \cite{Metz:2004ya} 
for a review.
Moreover, model results help to sharpen our physical intuition on these
novel objects, and can be used to make estimates for planned experiments.
Another aspect is that, thanks to the far simpler dynamics as compared to QCD, 
one may find relations among the different TMDs in some 
\cite{Jakob:1997wg,Avakian:2008dz,Pasquini:2008ax,Efremov:2008mp,She:2009jq}
though not all \cite{Bacchetta:2008af,Meissner:2007rx} models.
As all TMDs are a priori independent structures, any such {\sl model relations}
among TMDs are not expected to hold in QCD. 

It is interesting to ask, however, whether such relations could neverless be 
satisfied in nature at least approximately. 
Recalling that the nucleon is characterized by 8 leading-twist \cite{Boer:1997nt} 
and 16 subleading-twist \cite{Goeke:2005hb} TMDs, such approximate relations 
could be valuable, for the interpretation of first data, or for estimates 
for new experiments \cite{Avakian:2007mv}.

In order to judge to which extent a particular {\sl model relation} among TMDs 
might be respected by nature, till we know the answer from experiment, it is
helpful to understand under which general conditions in a model this relation
holds. 
For example, suppose a model relation relies on the SU(6) spin-flavour symmetry 
of the nucleon wave-function. We know from experiment that the SU(6) symmetry 
concept is useful --- within certain limitations
\cite{Boffi:2009sh,Close:1988br}.
This implies that model relations based on SU(6) symmetry 
are respected in nature, if at all, at best within similar limitations.
It happens that all model relations among TMDs observed so 
far have been found in models based on SU(6) symmetry
\cite{Jakob:1997wg,Avakian:2008dz,Pasquini:2008ax}.

The purpose of this work is to study TMDs in the covariant model of the 
nucleon proposed in Ref.~\cite{Zavada:1996kp} which makes {\sl no use}
of SU(6) spin-flavour symmetry. Some of the results presented here
were discussed previously in \cite{Efremov:2008mp}.
In this model the intrinsic motion of partons inside the nucleon is described 
in terms of a covariant momentum distribution. The model was applied 
to the study of unpolarized and polarized parton distribution functions
accessible in DIS $f_1^a(x)$, $g_1^a(x)$ and $g_T^a(x)$,
and extended to compute the transversity distribution $h_1^a(x)$
\cite{Zavada:2001bq,Zavada:2002uz,Efremov:2004tz,Zavada:2006yz,Zavada:2007ww}.

In this work we generalize the approach 
\cite{Zavada:1996kp,Zavada:2001bq,Zavada:2002uz,Efremov:2004tz,Zavada:2006yz,Zavada:2007ww}
to the description of TMDs.
In particular, we focus on the so called T-even, leading twist TMDs,
and pay particular attention to the demostration of the consistency of the 
approach. We shall see that certain model relations among TMDs hold 
{\sl even without} invoking SU(6) symmetry.

This note is organized as follows. 
In Sec.~\ref{Sec-2:TMDs} we briefly introduce TMDs. 
In Sec.~\ref{Sec-3:model} we introduce the model, and review previous works.
In Sec.~\ref{Sec-4:calculation} we generalize the approach to 
the description of TMDs, and in Sec.~\ref{Sec-5:consistency}
we demonstrate its consistency. In Sec.~\ref{Sec-6:relations}
we discuss the model relations among the polarized T-even TMDs. 
In Sec.~\ref{Sec-7:non-rel-limit} we apply the approach
to a study of TMDs in the non-relativistic limit,
before we summarize and conclude in Sec.~\ref{Sec-8:conclusions}.
The Appendices contain details of the calculations, and supplementory
results.

%\newpage

%===================  SECTION 2 ======================================
\section{TMDs}
\label{Sec-2:TMDs}

In this Section we introduce and define briefly TMDs.
With the use of light-cone coordinates, $a^\pm=(a^0\pm a^1)/\sqrt{2}$, 
TMDs are defined in terms of light-front correlators as
\be\label{Eq:correlator}
    \phi(x,\vec{p}_T)_{ij}=\int\frac{\di z^-\di^2\vec{z}_T}{(2\pi)^3}\;e^{ipz}
    \; \la N(P,S)|\bar\psi_j(0)\,{\cal W}(0,z,\mbox{path})\,\psi_i(z)|N(P,S)\ra
    \biggl|_{z^+=0,\,p^+ = xP^+} \;.
    \ee
In SIDIS the singled-out space-direction is along the momentum of the 
hard virtual photon $q^\mu=(q^0,q^1,0,0)$, and transverse vectors like 
$\vec{p}_T$ are perpendicular to it. The path of the Wilson-link depends on the 
process \cite{Collins:2002kn,Belitsky:2002sm}. 
In the nucleon rest frame the polarization vector is 
$S=(0,-S_L,\vec{S}_T)$ with $S_L^2+\vec{S}_T^2=1$.
The negative sign in front of $S_L$ is because by convention 
\cite{Bacchetta:2006tn} the nucleon has positive helicity, i.e.\ 
$S_L>0$, if it moves towards the virtual photon.

The information content of the correlator (\ref{Eq:correlator}) is 
summarized by eight leading-twist TMDs \cite{Boer:1997nt}, that can 
be projected out from the correlator (\ref{Eq:correlator}) as follows
(for convenience we will often suppress flavour indices)
\ba
    \frac12\;{\rm tr}\biggl[\gamma^+ \;\phi(x,\vec{p}_T)\biggr]         &=&
    \hspace{5mm}f_1 -\frac{\varepsilon^{jk}p_T^j S_T^k}{M_N}\,f_{1T}^\perp
    \label{Eq:TMD-pdfs-I}\\
    \frac12\;{\rm tr}\biggl[\gamma^+\gamma_5 \;\phi(x,\vec{p}_T)\biggr] &=&
    S_L\,g_1 + \frac{\vec{p}_T\cdot\vec{S}_T}{M_N}\,g_{1T}^\perp
    \label{Eq:TMD-pdfs-II}\\
    \frac12\;{\rm tr}\biggl[i\sigma^{j+}\gamma_5 \;\phi(x,\vec{p}_T)\biggr] &=&
    S_T^j\,h_1  + S_L\,\frac{p_T^j}{M_N}\,h_{1L}^\perp +
    \frac{(p_T^j p_T^k-\frac12\,\vec{p}_T^{\:2}\delta^{jk})S_T^k}{M_N^2}\,
    h_{1T}^\perp + \frac{\varepsilon^{jk}p_T^k}{M_N}\,h_1^\perp \;,
    \label{Eq:TMD-pdfs-III}
\ea
where the space-indices $j,k$ refer to the plane transverse with respect
to the light-cone, and $\varepsilon^{32} = - \varepsilon^{23} = 1$ and 
zero else (which is consistent with $\varepsilon^{0123}=1$).
Integrating out transverse momenta in the correlator (\ref{Eq:correlator})
leads to the three 'usual' parton distributions known from collinear kinematics
$j_1^a(x) = \int\di^2\vec{p}_T \, j_1^a(x,\vec{p}_T^{\:2})$ with $j=f,\,g,\,h$
\cite{Ralston:1979ys,Jaffe:1991ra}. Dirac-structures other than that in
Eqs.~(\ref{Eq:TMD-pdfs-I},~\ref{Eq:TMD-pdfs-II},~\ref{Eq:TMD-pdfs-III})
lead to subleading-twist terms \cite{Goeke:2005hb,Bacchetta:2006tn}.

%\newpage
%===================  SECTION 3 ======================================
\section{The covariant model of the nucleon, and its applicability}
\label{Sec-3:model}

In this Section we first briefly introduce the model, and sketch the 
calculation of the 'collinear' parton distribution functions done so far,
namely $f_1^a(x)$, $g_1^a(x)$, $g_T^a(x)$ and $h_1^a(x)$. 
Then we discuss the applicability of the approach  
to the calculation of TMDs which will be done in 
Sec.~\ref{Sec-4:calculation}. 

The starting point for the calculation of the chirally even 
functions accessible in DIS, $f_1^a(x)$, $g_1^a(x)$, $g_T^a(x)$
\cite{Zavada:1996kp,Zavada:2001bq,Zavada:2002uz}, is the hadronic tensor.
The latter is evaluated in the Bjorken-limit, i.e.\ in the limit that the 
four-momentum transfer $q^\mu$ from the lepton beam to the nucleon with 
momentum $P$ is such that $Q^2=-q^2$ and $Pq\to\infty$ while $x=Q^2/(2Pq)$ 
is fixed.
In the model it is assumed that unpolarized DIS can be described
as the incoherent sum of the scattering of electrons off 
non-interacting quarks, whose momentum distributions inside the nucleon 
are given in terms of the scalar function $G(pP/M)$. 
Here $p$ and $P$ are the momenta of the quark and nucleon, 
and $M$ is the nucleon mass.
Though all expressions can always be formulated in a manifestly
covariant way, it is convenient to work in the nucleon rest-frame,
where the momentum distribution becomes  $G(p^0)$ with 
$p^0=\sqrt{\vec{p}^{\;2}+m^2}$. Here $m$ denotes the quark mass.
Clearly, the distribution of the quark momenta in the nucleon 
rest frame is rotationally symmetric.

Applying these ideas to the description of the symmetric part of the 
hadronic tensor has shown that in the model the Callan-Gross relation 
among the unpolarized structure functions holds exactly, and the 
unpolarized parton distribution function is given by \cite{Zavada:1996kp}
(notice that $G(p^0)$ depends on flavour, which we suppress for brevity)
\be\label{Eq:f1-model}
 	 f_1^q(x) = \int\frac{\di^3 p}{p^0}\;G(p^0)\;
		    \delta\left(\frac{p^0-p^1}{M}-x\right)\;(p^0-p^1) \;.
\ee

Next we review the calculation of $g_1^q(x)$ and $g_T^q(x)$ 
appearing in the anti-symmetric part of the hadronic tensor. 
For a single quark the latter would be given by 
$W_{\alpha\beta}^{A,q}=m\epsilon_{\alpha\beta\mu\nu}q^\mu w^\nu$,
where $q^\mu$ is the momentum transfer from the electron in DIS,
and $w^\nu$ denotes the polarization vector of the quark.
In the model --- assuming the covariant distribution of polarized 
quarks to be given by $H(pP/M)$ --- the anti-symmetric part of the
hadronic tensor of the nucleon is given by
\be\label{Eq:W-mu-nu}
         W_{\alpha\beta}^{A} = \int\frac{\di^3 p}{Pp/M}\;
         H(pP/M)\delta((p+q)^2-m^2) W_{\alpha\beta}^{A,q}
         \stackrel{\rm Bj}{=} 
         \frac{m}{2Pq}\epsilon_{\alpha\beta\mu\nu}q^\mu
         \int\frac{\di^3 p}{p^0}\;H(p^0)
         \delta\biggl(\frac{p^0-p^1}{M}-x\biggr) w^\nu \;.
\ee
The second expression in (\ref{Eq:W-mu-nu}) is given in the nucleon 
rest frame choosing $q^\mu=(q^0,q^1,0,0)$ and holds in the Bjorken limit 
(more precisely: here and in the following for the steps marked by 
'Bj' the condition $Q^2\gg 4M^2x^2$ is essential).

Notice that the covariant distribution of polarized 
quarks can be expressed as $H(p^0)=G^+(p^0)-G^-(p^0)$, where the indizes
$(\pm)$ refer to the respective quark polarizations which are parallel (+) or 
anti-parallel ($-$) to the nucleon spin.
In this notation the covariant distribution of unpolarized quarks in
(\ref{Eq:f1-model}) is $G(p^0)=G^+(p^0)+G^-(p^0)$.

The most general expression for the covariant quark polarization 
vector \cite{Zavada:2001bq} is given by
\be\label{Eq:quark-pol-vec}
         w^\mu = - \frac{pS}{pP+mM}\,P^\mu + S^\mu 
                 - \frac{M}{m}\,\frac{pS}{pP+mM}\,p^\mu
\ee
where $S^\mu$ denotes the nucleon polarization vector given
in the nucleon rest frame by $S^\mu=(0,\vec{S})$ with $|\vec{S}|=1$. 
The evaluation of Eqs.~(\ref{Eq:W-mu-nu},~\ref{Eq:quark-pol-vec}) 
and comparison to the general Lorentz-decomposition of
$W_{\alpha\beta}^{A}$, namely
\be\label{Eq:W-mu-nu-gen}
         W_{\alpha\beta}^{A} = \epsilon_{\alpha\beta\mu\nu}q^\mu
         \biggl(\frac{S^\nu}{Pq}\;g_1 + 
         \frac{(Pq)S^\nu-(Sq)P^\nu}{(Pq)^2}\;g_2\biggr)\, ,
\ee
yield the following results for $g_1^q(x)$ and $g_T^q(x) = g_1^q(x) + g_2^q(x)$
\cite{Zavada:2001bq,Zavada:2002uz}
\ba
	g_1^q(x) &=& 
	\int\frac{\di^3 p}{p^0}\;H(p^0)\;\delta\left(\frac{p^0-p^1}{M}-x\right)
	\biggl[p^0-p^1 -\frac{\vec{p}_T^{\:2}}{p^0+m}\biggr] \; ,
        \label{Eq:g1-in-model}\\
	g_T^q(x) &=&
	\int\frac{\di^3 p}{p^0}\;H(p^0)\;\delta\left(\frac{p^0-p^1}{M}-x\right)
	\biggl[m+\frac{\vec{p}_T^{\:2}}{2(p^0+m)}\biggr]\label{Eq:gT-in-model}
        \;.
\ea
In the model the Burkhardt-Cottingham sum rule \cite{Burkhardt:1970ti} 
is satisfied. When neglecting terms proportional to $m$ 
also the Efremov-Leader-Teryaev sum rule \cite{Efremov:1996hd} holds, 
while $g_T^q(x)$ is given by the Wandzura-Wilczek (WW) 
approximation~\cite{Wandzura:1977qf} 
\be\label{Eq:WW}
        g_T^q(x)\stackrel{\rm WW}{=}\int^1_x \frac{\di y}{y} \; g_1^q(y)
        + {\cal O}\biggl(\frac{m}{M}\biggr)
\ee
as was proven in \cite{Zavada:2001bq,Zavada:2002uz}.
Notice that in QCD what is neglected are not only mass terms but
also pure twist-3 terms \cite{Wandzura:1977qf}.
That in the model such pure twist-3 ('interaction dependent') terms 
in $g_T(x)$ are absent, is {\sl consistent}
because in our approach the quarks are assumed to be free.

The chirally odd transversity distribution function cannot be accessed 
through the hadronic tensor and DIS. For theoretical purposes, however, 
one may consider the auxiliary polarized process described by the 
interference of a vector and a scalar current. On the quark level
this interference is described by  
$T_\mu^q=\epsilon_{\alpha\beta\lambda\nu} p^\beta q^\lambda w^\nu$
from which one obtains --- in analogy to the procedure in
Eq.~(\ref{Eq:W-mu-nu}) --- the following expression for the nucleon
\be\label{Eq:aux-vector}
        T_\alpha = \int\frac{\di^3 p}{Pp/M}\;
         H(pP/M)\delta((p+q)^2-m^2) T_\alpha^q
         \stackrel{\rm Bj}{=} 
         \frac{1}{2Pq}\epsilon_{\alpha\beta\lambda\nu}q^\lambda
         \int\frac{\di^3 p}{p^0}\;H(p^0)
         \delta\biggl(\frac{p^0-p^1}{M}-x\biggr) p^\beta w^\nu \;.
\ee
The general Lorentz-decomposition in this case reads
($j=2,3$ is a 'transverse index' with respect to $q$ and $P$)
\be\label{Eq:aux-vector-gen}
         2M(-1)\epsilon^{j\alpha}T_\alpha = S_T^j \,h_1^q(x)
\ee
one obtains after evaluating (\ref{Eq:aux-vector}) with 
(\ref{Eq:quark-pol-vec}) the following result for the transversity 
distribution function \cite{Efremov:2004tz}
\be\label{Eq:h1-in-model}
        h_1(x) = \int\frac{\di^3 p}{p^0}\;H(p^0)\;
	\delta\left(\frac{p^0-p^1}{M}-x\right)
	\biggl[p^0-p^1 -\frac{\vec{p}_T^{\:2}}{2(p^0+m)}\biggr] \;.
\ee

We remark that in 
Eqs.~(\ref{Eq:g1-in-model},~\ref{Eq:gT-in-model},~\ref{Eq:h1-in-model})
we did not distinguish momentum distributions in differently polarized 
nucleons. 
In general one might suspect the covariant distributions to be different.
In QCD, if nothing else, different
evolution properties clearly distinguish chirally even vs.\ odd, and
twist-2 vs.\ twist-3 distribution functions.
In the model, however, it is natural to assume the distributions in 
longitudinally and transversely polarized nucleons to be equal. In order
to understand that this assumption is indeed natural,
we recall that the approach is covariant. Therefore one may go to the 
nucleon rest frame, where it certainly makes no difference whether the 
quarks in the nucleon are polarized longitudinally or transversely 
(with respect to the space component of the four-vector $q$ in 
DIS or SIDIS).\footnote{
  Since $q^\mu$ is space-like $q^2 < 0$, its space-component $\vec{q}$ is 
  non-zero in any frame, and always provides an axis for the quantization
  of the nucleon spin.}
Since in this model the quarks are non-interacting,
it does not matter how they are polarized --- because, for example, 
there are also no spin-orbit- or spin-spin-interactions.
(Sec.~\ref{Subsec:exact-relation-in-QCD} will show
that the covariant distribution in various polarized TMDs must be equal,
in order to comply with QCD.)

However, this by no means implies that the parton distributions
describing longitudinally and transversely polarized quarks,
$g_1(x)$ and $h_1(x)$, are equal. They are, in fact, rather
different even if described in terms of the same covariant
momentum distribution $H(p^0)$ \cite{Efremov:2004tz}.
By introducing adequate normalizations (as dictated, e.g., by 
SU(6) symmetry) one could furthermore relate the polarized covariant 
distribution function $H(p^0)$ to the unpolarized one $G(p^0)$.
This is, however, a severe restriction and simplification 
of the model, which we do not need in general.

When extending the approach below to the description of TMDs
it is important to keep in mind the following point.
The QCD definition of a parton distribution function includes
a Wilson line, which in DIS describes the interaction of the struck
nucleon with the target remnant. It is possible to find a gauge 
in which the  Wilson line drops out, and the partons seem 'non-interacting'
--- an idea eventually underlying the parton model in general, and the
approach of Refs.~\cite{Zavada:2001bq,Zavada:2002uz,Zavada:2006yz,Zavada:2007ww,Efremov:2004tz}
in particular.
When dealing with TMDs, however, the Wilson line cannot be
'gauged away' \cite{Brodsky:2002cx,Collins:2002kn,Belitsky:2002sm}.

In the present framework we have no tool to include the effects
of the Wilson line, and therefore the description of the so-called 
'naively T-odd' parton distribution functions, the Sivers function 
$f_{1T}^\perp$ and the Boer-Mulders function $h_1^\perp$, is beyond
the scope of the approach. These TMDs crucially rely on the
initial- and final-state-ineractions encoded in the Wilson line
\cite{Brodsky:2002cx,Collins:2002kn}, and are expected to be absent
in our framework.

Finally, we remark that the model is 'opposite' to the Gaussian ansatz 
for TMDs in the following sense. In the Gaussian ansatz one assumes 
the extreme situation that distribution of longitudinal momentum
(i.e.\ $x$-dependence) and the distribution of $p_T$ are decoupled.
For example, one has 
$f_1(x,p_T^2) = f_1(x)\,\exp(-p_T^2/\la p_T^2\ra) / (\pi \la p_T^2\ra)$.
It is even possible to ``switch off'' $p_T$-effects: in the limit 
$\la p_T^2\ra\to 0$ one has $f_1(x,p_T^2) = f_1(x)\,\delta^{(2)}(\vec{p}_T)$.
In contrast to this in the present model the longitudinal and transverse
motions are coupled ``maximally''. It is not possible to ``switch off''
$p_T$-effects. As a consequence one has, e.g., interesting implications
for the quark orbital motion \cite{Zavada:2007ww}.

%\newpage
%===================  SECTION 4 ======================================
\section{Extension of the approach to TMDs}
\label{Sec-4:calculation}

In this Section we extend the approach to the description of TMDs. Since 
none of the new TMDs in Eqs.~(\ref{Eq:TMD-pdfs-I}--\ref{Eq:TMD-pdfs-III})
is accessible directly via the hadronic tensor in DIS or via the auxiliary 
process explored for the calculation of $h_1^q(x)$, we need to establish 
a more general relation in the model to the correlators
(\ref{Eq:TMD-pdfs-I}--\ref{Eq:TMD-pdfs-III}). 
For that we observe that the model expressions
for the anti-symmetric part of the hadronic tensor (\ref{Eq:W-mu-nu})
or the auxiliary current (\ref{Eq:aux-vector}) include integration
over $\di^3 p = \di p^1 \di^2p_T$ with $\di^2 p_T\equiv \di p_2\di p_3$.
In the following we will explore the consequences of what happens if one 
{\sl does not integrate out} transverse momenta in these expressions,
and demonstrate the consistency of this approach.

The 'integrated' symmetric part of the hadronic tensor, to which 
$f_1^q$ is related, was studied in Ref.~\cite{Zavada:1996kp}. The study 
of its 'unintegrated' version would give model results for $f_1^q$ and
the T-odd $f_{1T}^{\perp q}$, as revealed by the correlator in 
Eq.~(\ref{Eq:TMD-pdfs-I}).
However, the description of the Sivers function is beyond 
the scope of our approach, see Sec.~\ref{Sec-3:model}, and we therefore 
start with the more interesting case of the correlator (\ref{Eq:TMD-pdfs-I}) 
which describes $g_1^q(x,p_T)$ and the T-even TMD $g_{1T}^q(x,p_T)$.
(We shall come back to $f_1^q$ at the end of the next Section.)

In order to access the information contained in the correlator 
(\ref{Eq:TMD-pdfs-II}) we consider the transverse space components
($j,\,k=2,\,3$) of the 'unintegrated' anti-symmetric part of the hadronic 
tensor in Eq.~(\ref{Eq:W-mu-nu}). We work in the nucleon rest frame with 
the nucleon polarization vector as introduced in the sequence of 
Eq.~(\ref{Eq:correlator}) and choose $q^\mu = (q^0,q^1,0,0)$. 
Then, using (\ref{Eq:quark-pol-vec}) we obtain in the Bjorken-limit
\be\label{Eq:W-mu-nu-unintegrated}
        2MW_{jk}^A(x,\vec{p}_T) \stackrel{\rm Bj}{=}\;\epsilon_{jk} 
        \int\frac{\di p^1}{p^0}\;H(p^0)\,
        \delta\left(\frac{p^0-p^1}{M}-x\right)\Biggr\{
         -S_L\,\biggl(p^0-p^1-\frac{\vec{p}^{\;2}}{p^0+m}\biggr)
        -\frac{\vec{p}_T\vec{S}_T}{2}\;\frac{p^0-p^1}{p^0+m}
        \Biggr\} \;.
\ee
We recognize two contributions in (\ref{Eq:W-mu-nu-unintegrated}),
one proportional to the longitudinal nucleon spin component $S_L$
and one proportional to the projection of the nucleon spin on the transverse
parton momentum. These contributions coincide exactly with the decomposition
of the correlator in Eq.~(\ref{Eq:TMD-pdfs-II}). 
Thus, from the comparison of the coefficients we obtain
\ba
	 g_1^q(x,p_T) &=& \int\frac{\di p^1}{p^0}\;H(p^0)\;
	\delta\left(\frac{p^0-p^1}{M}-x\right)
	\biggl[\,p^0-p^1 -\frac{\vec{p}_T^{\:2}}{p^0+m}\biggr] \;,
        \label{Eq:g1-unintegrated}\\
	 g_{1T}^{\perp q}(x,p_T) &=& \int\frac{\di p^1}{p^0}\;H(p^0)\;
	\delta\left(\frac{p^0-p^1}{M}-x\right)
	\biggl[\,M\;\frac{p^0-p^1+m}{p^0+m}\biggr]\;.
        \label{Eq:g1Tperp-unintegrated}
\ea
Notice that there is no arbitrariness concerning an overall prefactor,
because in the integrated case we reproduce
$2MW_{jk}^A(x) = -\epsilon_{jk} S_L g_1^q(x)$ in agreement with the
general Lorentz-decomposition for the transverse components of the
anti-symmetric part of the hadronic tensor. The other components of 
$W_{\mu\nu}^A$ describe subleading twist structures, for example 
$g_T^q(x)$ in the integrated case, which we do not consider
in this work.

Now we wish to access the information content described in the
chirally odd correlator (\ref{Eq:TMD-pdfs-III}). 
For that we consider the 'unintegrated' version of the auxiliary
current $T_\alpha$ in Eq.~(\ref{Eq:aux-vector}). 
For $T_\alpha(x,\vec{p}_T)$ contracted with $\epsilon^{j\alpha}$ 
we obtain:
\be
	2M\,(-1)\varepsilon^{j\alpha}T_\alpha(x,\vec{p}_T) 
	= 
	\int\frac{\di p^1}{p^0}\;H(p^0)\;\delta\left(\frac{p^0-p^1}{M}-x\right)
	\Biggl\{S_T^j (p^0-p^1)
        - S_L p_T^j \;\frac{p^0-p^1+m}{p^0+m}
        - p_T^j\;\frac{\vec{S}_T\vec{p}_T}{p^0+m}
	\Biggr\}
  \, .\label{Eq:aux-vector-unintegrated} \ee
In order to easier compare to (\ref{Eq:TMD-pdfs-III}) we rewrite 
the decomposition of that correlator as often done 
\cite{Mulders:1995dh,Boer:1997nt} as follows
\ba\label{Eq:correlator-unintegrated-II}
    \frac12\;{\rm tr}
    \biggl[i\sigma^{j+}\gamma_5 \;\phi(x,\vec{p}_T)\biggr] 
    = S_T^j\,
    \underbrace{
      \Biggl(h_1^q-\frac{\vec{p}_T^{\:2}}{2M_N^2}\,h_{1T}^{\perp q}\Biggr)}_{
      \displaystyle h_{1T}^q}
    - S_L\,\frac{p_T^j}{M_N}\,h_{1L}^{\perp q}
    + \frac{p_T^j (\vec{p}_T\vec{S}_T)}{M_N^2}\,h_{1T}^\perp 
    + \frac{\varepsilon^{jk}p_T^k}{M_N}\,h_1^{\perp q} \;,
\ea
where we suppressed the arguments $x$, $p_T$ of the TMDs for brevity. 
By comparing the coefficients in 
Eqs.~(\ref{Eq:aux-vector-unintegrated},~\ref{Eq:correlator-unintegrated-II}) 
we read off the following results:
\ba
	h_{1T}^q(x,p_T) &=& 
	\int\frac{\di p^1}{p^0}\;H(p^0)\;\delta\left(\frac{p^0-p^1}{M}-x\right)
	\biggl[p^0-p^1\biggr]\label{Eq:h1T}\\
	h_{1L}^{\perp q}(x,p_T) &=& 
	\int\frac{\di p^1}{p^0}\;H(p^0)\;\delta\left(\frac{p^0-p^1}{M}-x\right)
	\biggl[-M\biggl(1-\frac{p^1}{p^0+m}\biggr)\biggr]\label{Eq:h1Lperp}\\
	h_{1T}^{\perp q}(x,p_T) &=& 
	\int\frac{\di p^1}{p^0}\;H(p^0)\;\delta\left(\frac{p^0-p^1}{M}-x\right)
	\biggl[-\frac{M^2}{p^0+m}\biggr]\label{Eq:pretzelosity}\\
	h_1^{\perp q}(x,p_T) &=& 0 \;.\phantom{\int\frac{\di p^1}{2p^0}}
\ea

%===================  SECTION 5 ======================================
\section{Consistency of the approach}
\label{Sec-5:consistency}

It is necessary to demonstrate the consistency of our approach.
For that we remark first that by integrating the expression
for $g_1^q(x,p_T)$ in Eq.~(\ref{Eq:g1-unintegrated}) over 
transverse momenta we recover the model result for $g_1^q(x)$
derived in \cite{Zavada:2001bq} and quoted in Eq.~(\ref{Eq:g1-in-model}).
Next, by exploring the connection of transversity to the functions 
$h_{1T}(x,p_T)$ and $h_{1T}^\perp(x,p_T)$ 
\cite{Mulders:1995dh,Boer:1997nt} 
\be
	h_1(x,p_T) =  h_{1T}(x,p_T) + \frac{\vec{p}_T^{\:2}}{2M^2}\,
        h_{1T}^\perp(x,p_T)
        = 
	\int\frac{\di p^1}{p^0}\;H(p^0)\;\delta\left(\frac{p^0-p^1}{M}-x\right)
	\biggl[p^0-p^1 -\frac{\vec{p}_T^{\:2}}{2(p^0+m)}\biggr]
        \label{Eq:transversity}\ee
and integrating over transverse momenta we recover the correct result 
for transversity derived in \cite{Efremov:2004tz} and quoted above in 
Eq.~(\ref{Eq:h1-in-model}). This means that the Lorentz-decomposition 
of the structure (\ref{Eq:aux-vector-unintegrated}) in the model is 
consistent with the Lorentz-decomposition of the correlator 
in Eqs.~(\ref{Eq:TMD-pdfs-III},~\ref{Eq:correlator-unintegrated-II}).
Since the model is covariant, this is an expected feature.
We finally notice that there is no polarization-independent term 
in (\ref{Eq:aux-vector-unintegrated}) meaning the absence of the
Boer-Mulders function --- as expected in the present framework, 
see Sec.~\ref{Sec-3:model}.

We observe that the results for $g_1^q(x,p_T)$ and $h_1^q(x,p_T)$
derived here could have been simply 'guessed' from the results for 
the integrated functions $g_1^q(x)$ and $h_1^q(x)$ by 'skipping'
the integration over $\di^2p_T$. This is by no means trivial,
because in the respective TMDs there could have been structures 
--- e.g., of the type $F(p^0,p^1)(p_2^2-p_3^2)$ with some function 
$F(p^0,p^1)$ making the integrals converging --- which would
drop out in the expression for the integrated functions due
to rotational symmetry in the transverse plane.
On the other hand, in the present approach described by a free 
Hamiltonian (that commutes with the momentum operator) the model
expressions for TMDs could be written as nucleon expectation values of 
certain polynomials of the momentum operator. Then it is clear that such 
'multipol-terms' in TMDs are forbidden by the Wigner-Eckart theorem.
After these considerations we conclude that the model expression
for the unintegrated unpolarized distribution function is given by
\be\label{Eq:f1-unintgrated}
 	 f_1^q(x,p_T) = \int\frac{\di p^1}{p^0}\;G(p^0)\;
		    \delta\left(\frac{p^0-p^1}{M}-x\right)\;(p^0-p^1) \;.
\ee

As a next important consistency check let us test whether the model results 
satisfy positivity constraints, and for that we need the expression for 
the unpolarized distribution function (\ref{Eq:f1-unintgrated}), since 
the inequalities for TMDs we wish to verify read \cite{Bacchetta:1999kz}
\ba
          |h_1^q(x,p_T)| &\le& 
          \frac12\biggl[f_1^q(x,p_T)+g_1^q(x,p_T)\biggr]\label{Ineq:I}\\
          |h_{1T}^{\perp(1)q}(x,p_T)| &\le&
          \frac12\biggl[f_1^q(x,p_T)-g_1^q(x,p_T)\biggr]\label{Ineq:II}\\
          g_{1T}^{\perp(1)q}(x,p_T)^2+f_{1T}^{\perp(1)q}(x,p_T)^2 &\le&
          \frac{\vec{p}_T^{\:2}}{4M^2}
          \biggl[f_1^q(x,p_T)^2-g_1^q(x,p_T)^2\biggr]\label{Ineq:III}\\
          h_{1L}^{\perp(1)q}(x,p_T)^2+h_1^{\perp(1)q}(x,p_T)^2 &\le&
          \frac{\vec{p}_T^{\:2}}{4M^2}
          \biggl[f_1^q(x,p_T)^2-g_1^q(x,p_T)^2\biggr] \label{Ineq:IV}
\ea
where the 'unintegrated' transverse moment of a TMD is defined as
\be\label{Eq:def-transv-mom}
        j_1^{(1)}(x,p_T) = \frac{\vec{p}_T^{\:2}}{2M^2}\;j_1(x,p_T)\;.
\ee
Notice that the T-odd functions $f_{1T}^{\perp(1)q}(x,p_T)$ and
$h_1^{\perp(1)q}(x,p_T)$ are absent in our approach.
A direct test of the inequalities is actually difficult because 
different covariant functions $G(p^0)$, $H(p^0)$ apprear in the TMDs.
%(and the $H(p^0)$ in the various polarized TMDs could in general also 
%be different). 
One way to proceed is to assume SU(6) spin-flavour 
symmetry of the nucleon wave function. If one assumes SU(6)
symmetry, the inequalities (\ref{Ineq:I}--\ref{Ineq:IV}) are manifestly 
satisfied in our framework, see App.~\ref{App:inequalities}.
If one does not, the positivity conditions (\ref{Ineq:I}--\ref{Ineq:IV}) 
are 'translated' into certain constraints among the covariant momentum 
distributions $G(p^0)$ and %(in general, the various) 
$H(p^0)$, see Ref.~\cite{Efremov:2004tz}, where the $p_T$-integrated 
version of (\ref{Ineq:I}) known as Soffer bound \cite{Soffer:1994ww} 
was discussed in this way.

We conclude this Section with the observation that so far our
approach satisfied all imposed consistency checks. 
Further consistency tests will be provided in the next Section,
where we shall see that our approach satisfies certain {\sl exact relations}
as well as {\sl model relations} among different TMDs found also in other 
relativistic quark models. 

%\newpage
%===================  SECTION 6 ======================================
\section{Relations in the model among TMDs}
\label{Sec-6:relations}

It has to be stressed that in QCD all TMDs are independent structures, 
and it is not possible to find exact relations among them that would allow
to express one TMD in terms of other TMDs.
However, especially in models without gauge field degrees of freedom 
\cite{Jakob:1997wg,Avakian:2008dz,Pasquini:2008ax}, it might be possible 
to find relations among different (T-even) TMDs.
Such model relations are of interest by themselves, and might be
supported by data within certain 'model accuracies'. It would be 
interesting to know the general conditions a quark-model must satisfy 
in order to fullfill such relations.

In order to recognize more easily the relations among the different T-even 
TMDs let us introduce the following compact notation for the measures
\ba
        \{\di\tilde{p}^1\} &\equiv& \frac{\di p^1}{p^0}\;\frac{G(p^0)}{p^0+m}\;
        \delta\left(\frac{p^0-p^1}{M}-x\right) \label{Eq:measure-f1} \\
        \{\di p^1\} &\equiv& \frac{\di p^1}{p^0}\;\frac{H(p^0)}{p^0+m}\;
        \delta\left(\frac{p^0-p^1}{M}-x\right) \label{Eq:measure-etc} .
\ea
The measure (\ref{Eq:measure-f1}) is positive definite, while the sign
of the measure (\ref{Eq:measure-etc}) depends on the sign of $H(p^0)$.
Then the various TMDs can be written as follows
\ba
 	 f_1^q(x,p_T) &=& \int\{\di\tilde{p}^1\}\;
         \biggl[\;(p^0+m)\;xM\biggr] \label{Eq:comp-f1}\\
	 g_1^q(x,p_T) &=& \int\{\di p^1\}\;
         \biggl[\;(p^0+m)\;xM-\vec{p}_T^{\:2}\biggr] \label{Eq:comp-g1}\\
         h_1^q(x,p_T) &=& \int\{\di p^1\}\;
         \biggl[\;(p^0+m)\;xM-\frac12\vec{p}_T^{\:2}\biggr] 
         = \int\{\di p^1\}\;\biggl[\;\frac12\,(xM+m)^2\biggr]
         \label{Eq:comp-h1}\\
	 g_{1T}^{\perp q}(x,p_T) &=& \int\{\di p^1\}\;
         \biggl[+\,M\;(xM+m)\biggr]\label{Eq:comp-g1Tperp}\\
         h_{1L}^{\perp q}(x,p_T) &=& \int\{\di p^1\}\;
         \biggl[-\,M\;(xM+m) \biggr]\label{Eq:comp-h1Lperp}\\
         h_{1T}^{\perp q}(x,p_T) &=& \int\{\di p^1\}\;
         \biggl[-\,M^2\biggr]\label{Eq:comp-h1Tperp}
\ea
where we remind that $p^0=\sqrt{p_1^2+\vec{p}_T^{\;2}+m^2}$ 
and $p^0-p^1 = xM$ due to the delta-function in the measures. 
When deriving the second equality in (\ref{Eq:comp-h1}) one may make use of 
the identity $\vec{p}_T^{\:2}=xM(p^0+p^1)-m^2$ valid due to the delta-function.
From the expressions (\ref{Eq:comp-g1}--\ref{Eq:comp-h1Tperp}) we can read
off numerous {\sl model} relations among polarized TMDs, which we shall
discuss in the following. 

Notice that none of the relations discussed in the following involves the 
unpolarized parton distribution function. Such relations are {\sl impossible} 
in our approach, simply because there is in general no way to connect the 
different covariant distributions $G(p^0)$ and $H(p^0)$.
If (and only if) one makes an additional model assumption, namely assumes
the SU(6) spin-flavour symmetry, then we obtain relations including 
$f_1^q(x,p_T)$ and other TMDs, see App.~\ref{App:relations-in-SU6}.

\subsection{An exact relation in QCD, and its consistent realization in the model}
\label{Subsec:exact-relation-in-QCD}

Let us first discuss an {\sl exact} relation which is valid in the model 
and which involves $g_T^q(x)$. (Later we shall discuss also several 
{\sl approximate} relations involving this twist-3 parton distribution 
function.)
For that we introduce the measure $\{\di^3 p\}=\{\di p^1\}\di^2p_T$
which allows us to rewrite the model expression (\ref{Eq:gT-in-model}) as 
\be\label{Eq:gT-in-model-II}
        g_T^q(x)=\int\{\di^3 p\}
        \biggl[\;m\,(p^0+m)+\frac12\,\vec{p}_T^{\:2}\biggr]\;.
\ee
From Eqs.~(\ref{Eq:comp-h1},~\ref{Eq:comp-g1Tperp},~\ref{Eq:gT-in-model-II})
we find that the following QCD relation \cite{Bacchetta:2006tn} is satisfied 
consequently in the model
\be\label{Eq:EOM-relation}
        x\,g_T^q(x)=
        g_{1T}^{\perp(1)q}(x)+\frac{m}{M}\,h_1^q(x)+
        \underbrace{x\,\widetilde{g}_T^q(x)}_{=0, \;\rm here!\hspace{-1cm}}\,.
\ee
This further demonstrates the consistency of our approach,
although $\tilde{g}_T^q(x)\neq 0$ in general, because in our model 
such pure twist-3 terms (quark-gluon-correlations) 
are consequently absent. We learn two further important lessons. 

First, the relation (\ref{Eq:EOM-relation}) crucially relies on the fact 
that $g_T^q$, $h_1^q(x)$, $g_{1T}^{\perp q}$ are described in terms of the 
{\sl same} covariant distribution function $H(p^0)$. In other words,
to be consistent with QCD, we actually have no choice but must work
with the same  covariant distribution function $H(p^0)$ 
for all polarized functions (c.f.\ the discussion in Sec.~\ref{Sec-3:model}). 
Second, we clearly see that the model
parameter $m$ is really to be identified with the current quark mass in QCD.

Let us remark that in QCD the relation (\ref{Eq:EOM-relation}) is valid
also in 'unintegrated version', i.e.\ with the TMDs not integrated over $p_T$.
Here we confine 
ourselves to the 'integrated relation' (\ref{Eq:EOM-relation}), 
as we have not derived the model expression for $g_T^q(x,p_T)$
(though, in view of the experience with $g_1^q$ and $h_1^q$, presumably it is
given by (\ref{Eq:gT-in-model},~\ref{Eq:gT-in-model-II}) with $p_T$-integration 
omitted and the 'unintegrated version' of (\ref{Eq:EOM-relation}) is valid, too).

\subsection{Exact relations among leading-twist TMDs in the model}

Next we focus on a class of relations among leading twist TMDs
which are exact in our approach, and can hold in models only.
By comparing the expressions in 
Eqs.~(\ref{Eq:comp-g1Tperp},~\ref{Eq:comp-h1Lperp})
we observe the following {\sl exact} relation in our model
\be\label{Eq:strong-relation-I}
        g_{1T}^{\perp q}(x,p_T) = - h_{1L}^{\perp q}(x,p_T) \;.
\ee
This relation was observed previously in the spectator model of 
Ref.~\cite{Jakob:1997wg} and the constituent model \cite{Pasquini:2008ax}.

Next, by comparing 
Eqs.~(\ref{Eq:comp-g1},~\ref{Eq:comp-h1},~\ref{Eq:comp-h1Tperp}) 
we see that the model results satisfy the relation
\be\label{Eq:measure-of-relativity}
	g_1^q(x,p_T) - h_1^q(x,p_T) = h_{1T}^{\perp(1)q}(x,p_T)\;.
\ee
This relation was first observed in the bag model \cite{Avakian:2008dz}
and it is also valid in the spectator model of Ref.~\cite{Jakob:1997wg}.
It was argued \cite{Avakian:2008dz} that (\ref{Eq:measure-of-relativity}) 
could be valid in a larger class of {\sl relativistic} models. It is thus 
gratifying to observe that subsequently (\ref{Eq:measure-of-relativity}) 
was confirmed in the relativistic constituent quark model 
\cite{Pasquini:2008ax} and now also in our approach.
Interestingly, the relation  (\ref{Eq:measure-of-relativity}) is not
supported in the spectator model version of Ref.~\cite{Bacchetta:2008af}.

Both quark model relations, 
Eqs.~(\ref{Eq:strong-relation-I},~\ref{Eq:measure-of-relativity}),
are not supported in models with gauge-field dregrees of freedom
\cite{Meissner:2007rx}.
This observation is in line with the expectation that even if the 
relations (\ref{Eq:strong-relation-I},~\ref{Eq:measure-of-relativity})
were valid in QCD at some scale (which, of course, does not need to be 
the case) they would be spoiled at any different scale by evolution 
effects that clearly discriminate chirally even and odd functions.

Finally, from 
Eqs.~(\ref{Eq:h1T},~\ref{Eq:h1Lperp},~\ref{Eq:pretzelosity})
we find the following remarkable exact relation 
\be\label{Eq:strong-relation-III}
        \frac12\biggl[\,h_{1L}^{\perp q}(x,p_T)\,\biggr]^2 = 
        - \;h_1^q(x,p_T)\;h_{1T}^{\perp q}(x,p_T) \;,
\ee
that was not observed before in literature to best of our 
knowledge and connects only chirally odd TMDs --- in contrast to 
(\ref{Eq:strong-relation-I},~\ref{Eq:measure-of-relativity}). This non-linear 
relation is not obeyed in the spectator model \cite{Jakob:1997wg}. 
Combining (\ref{Eq:strong-relation-I},~\ref{Eq:strong-relation-III}) we find 
\be\label{Eq:strong-relation-IV}
        \frac12\biggl[\,g_{1T}^{\perp q}(x,p_T)\,\biggr]^2 = 
        - \;h_1^q(x,p_T)\;h_{1T}^{\perp q}(x,p_T) \;,
\ee
which again mixes chirally odd and even TMDs
(though the product of two chirally odd objects 'conserves' chirality).

From Eqs.~(\ref{Eq:comp-g1},~\ref{Eq:comp-h1}) we find also relations
among the signs of TMDs, for example:
\ba\label{Eq:strong-relation-sign-1}
        {\rm sign}(h_1^q) &=& \phantom{-}\;{\rm sign}(g_1^q)\;,\\
   \label{Eq:strong-relation-sign-2}
        {\rm sign}(h_1^q) &=& -\;{\rm sign}(h_{1T}^{\perp q})\;.
\ea
Notice that (\ref{Eq:strong-relation-sign-2}) could be concluded as a 
corollary from the result (\ref{Eq:strong-relation-III}).
Also from Eqs.~(\ref{Eq:comp-g1},~\ref{Eq:comp-h1}), 
or from combining  (\ref{Eq:strong-relation-sign-1}) and 
(\ref{Eq:measure-of-relativity}), we find 
\be\label{Eq:strong-relation-VI}
        |h_1^q(x,p_T)| > |g_1^q(x,p_T)|  \;\;\;\mbox{for}\;\;p_T>0\;,
\ee
i.e.\  the modulus of the transversity distribution function 
is larger than that of the helicity distribution function.
This inequality survives integration over $p_T$ and $x$, and
we obtain for $g_T^q=\int\di x\,h_1^q(x)$ and $g_A^q=\int\di x\,g_1^q(x)$,
the tensor and axial charges, the relation
\be\label{Eq:strong-relation-VII}
        |g_T^q| > |g_A^q|  \;\;,
\ee
which was also observed in \cite{Efremov:2004tz}, in many other models 
\cite{Pasquini:2005dk,Schweitzer:2001sr,Barone:2001sp}, and in lattice QCD 
\cite{Gockeler:2006zu}.

Although some of these results have been obtained before in literature,
it is remarkable that in our approach all these relations follow 
{\sl without} assuming SU(6) spin-flavour symmetry.

\subsection{Relations in the model in the chiral limit}

From Eqs.~(\ref{Eq:comp-h1}--\ref{Eq:comp-h1Tperp}) one can find
further relations by observing that $h_1^q$, $g_{1T}^{\perp q}$, 
$h_{1L}^{\perp q}$, $h_{1T}^{\perp q}$ are functions of the type
${\rm const}\times(xM+m)^n\times\int\{\di p^1\}$, which looks 'so trivial'
only due to the compact notation introduced in Eq.~(\ref{Eq:measure-etc}).
Recall that $m$ is to be identified with the QCD current quark mass,
see Sec.~\ref{Subsec:exact-relation-in-QCD}, whose effects are expected 
to be negligible in deeply inelastic reactions. 
We can formulate those relations in the chiral limit, and obtain
\ba\label{Eq:chi-lim-1}
   	2h_1^q(x,p_T) + x^2\,h_{1T}^{\perp q}(x,p_T) =
        {\cal O}\biggl(\frac{m}{M}\biggr)\;,\\
   \label{Eq:chi-lim-2}
	2h_1^q(x,p_T) \,+\, x\,h_{1L}^{\perp q}(x,p_T) =
        {\cal O}\biggl(\frac{m}{M}\biggr)\;,\\
   \label{Eq:chi-lim-3}
   	h_{1L}^{\perp q}(x,p_T) - x\,h_{1T}^{\perp q}(x,p_T) =
        {\cal O}\biggl(\frac{m}{M}\biggr)\;.
\ea
Upon the use of (\ref{Eq:strong-relation-I}) one obtains 
relations similar to (\ref{Eq:chi-lim-2},~\ref{Eq:chi-lim-3})
but with $h_{1L}^{\perp q}$ replaced by $(-g_{1T}^{\perp q})$.

Remarkably, in the model it is possible to relate the transverse moments of
the chirally odd TMDs $h_{1L}^{\perp q}$ and $h_{1T}^{\perp q}$ to the 
chirally even twist-3 parton distribution function $g_T^q(x)$ as follows
\ba\label{Eq:chi-lim-4a}
  h_{1L}^{\perp(1)q}(x)+x\,g_T^q(x)={\cal O}\biggl(\frac{m}{M}\biggr)\;,
  \\
  \label{Eq:chi-lim-4b}
  h_{1T}^{\perp(1)}(x) + g_T(x) = {\cal O}\biggl(\frac{m}{M}\biggr)\;.
\ea
Notice that (\ref{Eq:chi-lim-4a}) follows also from combining
(\ref{Eq:EOM-relation}) and (\ref{Eq:strong-relation-I}).

\subsection{Wandzura-Wilczek (type) approximations}

We already mentioned the Wandzura-Wilczek approximation \cite{Wandzura:1977qf}
which allows to connect the twist-3 distribution function $g_T^q(x)$ and the
twist-2 distribution function $g_1^q(x)$, see Eq.~(\ref{Eq:WW}).
In the model this approximation requires the neglect of quark mass terms.
In QCD one has to neglect in addition pure twist-3 terms.

An important practical application is that  
Eqs.~(\ref{Eq:WW},~\ref{Eq:chi-lim-4b}) allow to express the unknown 
$h_{1T}^{\perp(1)q}(x)$ in terms of the well-known $g_1^q(x)$. 
In \cite{Efremov:2008mp} we made use of this relation in order to estimate
the transverse moment of pretzelosity, and used the results for 
predictions of SSAs in SIDIS.

In a similar way, i.e.\ upon the neglect of pure twist-3 and quark mass terms, 
one obtains further 'Wandzura-Wilczek-type approximations' \cite{Avakian:2007mv},
namely
\ba
    	g_{1T}^{\perp(1)q}(x) &\approx& 
        \;\;\;x\;\int_x^1\frac{\di y}{y\;\;}\,g_1^q(y)\;,
	\label{Eq:WW-approx-g1T}\\
    	h_{1L}^{\perp(1)q}(x) &\approx& 
        - x^2\!\int_x^1\frac{\di y}{y^2\;}\,h_1^q(y)\;.
	\label{Eq:WW-approx-h1L}
\ea
Both Wandzura-Wilczek-type approximations are valid in our approach upon 
the neglect of quark-mass terms. The validity of (\ref{Eq:WW-approx-g1T}) 
follows directly from Eqs.~(\ref{Eq:WW},~\ref{Eq:EOM-relation}).
The proof of the Wandzura-Wilczek-type approximation (\ref{Eq:WW-approx-h1L}) 
is given in App.~\ref{App:proof-of-WW-type-relation}.

\subsection{Transverse momentum dependence}
\label{Sec:pT-dependence}

The probably most exciting thing about TMDs is, of course, their $p_T$-dependence.
The power of the covariant approach with rotationally symmetric momentum 
distributions of quarks in the nucleon rest frame is based on the fact 
that this symmetry ultimately connects the distributions of transverse 
and longitudinal momenta. 
Some exciting consequences of this symmetry in the context of the spin
content were discussed in \cite{Zavada:2007ww}, and a detailed study of 
the effects of this symmetry in the context of TMDs is in preparation.

However, a couple of simple but already interesting conclusions on 
the 'mean transverse momenta' of TMDs can be drawn without modelling 
the covariant momentum distribution $H(p^0)$. Notice that all $p_T$-integrals 
given below are well-defined in our approach.

Let us introduce the notion of mean transverse momenta moments 
of a TMD $j_1$ as follows
\be
         \la \vec{p}_T^{\:2},j_1\ra = 
         \frac{\int\di x\int\di^2p_T\;\vec{p}_T^2\;j_1(x,p_T)}
              {\int\di x\int\di^2p_T\;             j_1(x,p_T)} \;,
\ee
and the $n$-th moment $p$-moment ($p=|\vec{p}|$) of the covariant 
distribution function is defined as follows
\be
         \la\la p^n\ra\ra = \int\di^3p\;p^n\,H(p^0)\;.
\ee
Then we obtain the following relations valid in the chiral limit
\ba
        \lim\limits_{m\to0}\la \vec{p}_T^{\:2},g_1^q\;\ra  =
        \lim\limits_{m\to0}\la \vec{p}_T^{\:2},h_1^q\ra  
        &=& \frac{2}{3}\;\frac{\la\la p^2\ra\ra}{\la\la 1\ra\ra}
        \label{Eq:pT2-chi-lim-1}\\
        \lim\limits_{m\to0}\la \vec{p}_T^{\:2},g_{1T}^{\perp q}\ra  =
        \lim\limits_{m\to0}\la \vec{p}_T^{\:2},h_{1L}^{\perp q}\ra  
        &=& \frac{2}{3}\;\frac{\la\la p \ra\ra}{\la\la p^{-1} \ra\ra}
        \label{Eq:pT2-chi-lim-2}\\
        \lim\limits_{m\to0}\la \vec{p}_T^{\:2},h_{1T}^{\perp q}\ra  
        &=& \frac{2}{3}\;\frac{\la\la 1\ra\ra}{\la\la p^{-2}\ra\ra}\;.
        \label{Eq:pT2-chi-lim-3}
\ea
It is instructive to learn that, although $g_1(x)$ and $h_1(x)$ are very
different in the model \cite{Efremov:2004tz}, their mean transverse
momenta coincide in the chiral limit.

%\newpage
%===================  SECTION 7 =====================================
\section{Non-relativistic limit}
\label{Sec-7:non-rel-limit}

The assumption of a non-relativistic dynamics of light quarks in the nucleon 
is not realistic. Nevertheless certain conclusions from this limit are very 
popular --- like, for example, the relation $h_1(x) = g_1(x)$ which has 
often been used in literature to obtain order of magnitude estimates 
for effects of transversity. It is therefore worth to study 
how this limit can be formulated in our framework.
This will yield non-relativistic limit results for TMDs.

In the strict non-relativistic limit, we have particle conservation 
and the nucleon consists of exactly $3$ (valence) quarks.
Then we deal with the dynamics of constituent ('valence') quarks,
whose momenta become negligible with respect to $m$, and whose binding
energy becomes negligible with respect to the nucleon mass such that
$M=3m$ up to relativistic corrections. The heavy constituent quarks
obey spin-flavour symmetry which is introduced in the context of
TMDs in App.~\ref{App:inequalities} in Eq.~(\ref{Eq-app:assume-SU6}).

The non-relativistic limit makes the following predictions for the
collinear parton distribution functions
\be\label{Eq:non-rel-limit}
	\limNonRel f_1^q(x) = N_q\,\delta\left(x-\frac13\right)\;,\;\;\;
	\limNonRel g_1^q(x) = 
	\limNonRel h_1^q(x) = P_q\,\delta\left(x-\frac13\right)\;.
\ee
If the momenta of quarks are not negligible with respect to their mass $m$, 
the $\delta$-functions in Eq.~(\ref{Eq:non-rel-limit}) are spread out. 
The normalizations $N_q$ and $P_q$ in Eq.~(\ref{Eq:non-rel-limit}) 
dictated by SU(6) spin-flavour symmetry are given in
Eq.~(\ref{Eq-app:assume-SU6}) in App.~\ref{App:inequalities}.
One can check that all sum rules are correctly reproduced:
\ba
  \int\di x\;   f_1^q(x) &=& N_q    \;\;\;\mbox{('normalization')}  
  \label{Eq:sum-rules}\\
  \sum_q\int\di x\;x\,f_1^q(x) &=& 1\;\;\;\mbox{(momentum sum rule)}\nonumber\\
  \sum_q\int\di x\;   g_1^q(x)  = 
  \sum_q\int\di x\;   h_1^q(x)  = 
         g_A^{(0)} = g_T^{(0)} &=& 1
  \;\;\;\mbox{('spin sum rule'/isoscalar tensor charge)} \nonumber\\
	\int\di x\;\biggl(g_1^u(x)-g_1^d(x)\biggr) =
	\int\di x\;\biggl(h_1^u(x)-h_1^d(x)\biggr) =  
  	g_A^{(3)}  = g_T^{(3)} &=& \frac53
  	\;\;\;\mbox{(Bjorken sum rule/isovector tensor charge).} \nonumber\ea

The model is formulated in a covariant way, which means that it also can 
be applied to the situation when the motion of quarks is assumed to be
non-relativistic, i.e.\ when $|\vec{p}|\ll m$ and 
$p_0=m\{1+{\cal O}(\vec{p}^{\:2}/m^2)\}$. 
In order to formulate the non-relativistic limit in our model,
we assume SU(6) symmetry and set the covariant momentum distributions in 
unpolarized or polarized nucleons equal,
i.e.\ we assume $G(p^0)\to N_qJ(\vec{p})$ and $H(p^0)\to P_qJ(\vec{p})$
with $\int\di^3 \vec{p}\;J(\vec{p}) = 1$.
Of course, $J(\vec{p})$ strictly speaking depends only on the modulus 
$|\vec{p}|$ (or $p^0=\sqrt{m^2+|\vec{p}|^2}$) due to the rotational symmetry
in the nucleon rest frame, but this notation is more convenient
for the following. 

In the non-relativistic limit one expects only small momenta $\vec{p}\to0$ to
be relevant in the integral over $\di^3\vec{p}$. Thus, it is natural to assume
\be\label{Eq:non-rel-lim-in-model}
	\limNonRel J(\vec{p}) \to \delta^{(3)}(\vec{p})
\ee
This is in some sense an 'axiom', and we have to verify that it yields to 
consistent results. For that we insert (\ref{Eq:non-rel-lim-in-model}) 
in Eq.~(\ref{Eq:f1-model}), and obtain (notice that $p^0\to m$ for 
$|\vec{p}|\ll m$)
\be\label{Eq:f1-in-model-non-rel-limit}
	\limNonRel f_1^q(x) = N_q \; x M \int\frac{\di^3 \vec{p}}{p^0}\;
	\delta^{(3)}(\vec{p}) \; \delta\left(\frac{p^0+p^1}{M}-x\right)
	= N_q  \underbrace{\;\frac{x M}{m}\;}_{=1} 
        \delta\left(\frac{m}{M}-x\right)
	= N_q  \delta\left(x-\frac{1}{3}\right)\;.
\ee

Thus, our prescription (\ref{Eq:non-rel-lim-in-model}) gives the correct
non-relativistic result. What do we obtain for the collinear polarized 
distribution functions? Let us insert (\ref{Eq:non-rel-lim-in-model}) 
in Eqs.~(\ref{Eq:g1-in-model},~\ref{Eq:gT-in-model},~\ref{Eq:h1-in-model})
We obtain
\be\label{Eq:non-rel-PDFs}
	\limNonRel g_1^q(x)  = \limNonRel g_T^q(x) = \limNonRel h_1^q(x)
	= P_q\,\delta\left(x-\frac13\right) \;.
\ee
Thus, the polarized distributions $g_1^q(x)$, $g_T^q(x)$, $h_1^q(x)$ become 
equal and correctly reproduce the non-relativistic result 
(\ref{Eq:non-rel-limit}). Thus, we see that our formulation of the
non-relativistic limit also yields correct results for the polarized
parton distribution functions. 

Let us now apply the non-relativistic limit to the description of
TMDs. We obtain
\ba
\limNonRel f_1^q(x,p_T) =  \label{Eq:non-rel-f1}
           N_q\;\delta\!\left(x-\frac{1}{N_c}\right)\;\delta^{(2)}(\vec{p}_T)\;,\\
\limNonRel g_1^q(x,p_T) =  \label{Eq:non-rel-g1}
           P_q\;\delta\!\left(x-\frac{1}{N_c}\right)\;\delta^{(2)}(\vec{p}_T)\;,\\
\limNonRel h_1^q(x,p_T) =  \label{Eq:non-rel-h1}
           P_q\;\delta\!\left(x-\frac{1}{N_c}\right)\;\delta^{(2)}(\vec{p}_T)\;,\\
\limNonRel g_{1T}^{\perp q}(x,p_T) =  \;\phantom{+}N_c\;\label{Eq:non-rel-g1Tperp}
 	   P_q\;\delta\!\left(x-\frac{1}{N_c}\right)\;\delta^{(2)}(\vec{p}_T)\;,\\
\limNonRel h_{1L}^{\perp q}(x,p_T) = \;-\,N_c\;\label{Eq:non-rel-h1Lperp}
 	   P_q\;\delta\!\left(x-\frac{1}{N_c}\right)\;\delta^{(2)}(\vec{p}_T)\;,\\
\limNonRel h_{1T}^{\perp q}(x,p_T) =  -\frac{N_c^2}{2}\;\label{Eq:non-rel-h1Tperp}
 	   P_q\;\delta\!\left(x-\frac{1}{N_c}\right) \;\delta^{(2)}(\vec{p}_T)\;.
\ea
In $g_{1T}^{\perp q}(x,p_T)$ and $h_{1L}^{\perp q}(x,p_T)$ the factors 
$\frac{M}{m}=N_c$ appear, while in $h_{1T}^{\perp q}(x,p_T)$ the factor 
$\frac{M^2}{2m^2}=\frac{N_c^2}{2}$ appears with $N_c=3$ colours.
These factors appear here somehow artificially because the nucleon mass was 
chosen in the Lorentz-decomposition of the correlators 
(\ref{Eq:TMD-pdfs-I}--\ref{Eq:TMD-pdfs-III}) to compensate 
the dimension of transverse momentum. Nevertheless, once one introduces the 
nucleon mass in this context (and sets the according 'units to measure' TMDs), 
the integrated functions $g_{1T}^{\perp q}(x)$,  $h_{1L}^{\perp q}(x)$,
$h_{1T}^{\perp q}(x)$ are larger then the parton distributions
$g_1^q(x)$ and  $h_1^q(x)$. It even happens that 
$|h_{1T}^{\perp q}(x)|>f_1^q(x)$ as also observed in other models
\cite{Avakian:2008dz,Pasquini:2008ax}. This is not in contradiction
with positivity which constrains only the transverse moments
of TMDs, see Eqs.~(\ref{Ineq:II}--\ref{Ineq:IV}).

From Eqs.~(\ref{Eq:non-rel-f1}-\ref{Eq:non-rel-h1Tperp}) we see
that the transverse moments of all TMDs vanish. In particular,
$h_{1T}^{\perp(1) q}(x)$ vanishes. This is consistent from the
point of view of the relation between helicity, transversity
and (the transverse moment of) pretzelosity, 
Eq.~(\ref{Eq:measure-of-relativity}), since in the non-relativistic limit, 
helicity and transversity distributions become equal (\ref{Eq:non-rel-PDFs}).

Thus, a non-zero transverse moment of pretzelosity \cite{Avakian:2008dz}
or any other TMD (as we learn here) can be considered to be a 'measure of
relativistic' effects in the nucleon. Clearly, any effect of TMDs would
dissapear from a cross section (or spin asymmetry). However, the TMDs 
themselves are all non-zero in the non-relativistic limit, see 
Eqs.~(\ref{Eq:non-rel-f1}-\ref{Eq:non-rel-h1Tperp}).

%===================  SECTION 8 =====================================
\section{Conclusions}
\label{Sec-8:conclusions}

We have generalized the covariant model developed in
Refs.~\cite{Zavada:1996kp,Zavada:2001bq,Zavada:2002uz,Efremov:2004tz,Zavada:2006yz,Zavada:2007ww}
to the description of T-even leading-twist TMDs in the nucleon. 
We have payed particular attention to the demonstration of the consistency 
of the extended approach.
For example, we have shown that it gives the familiar results for the 
'integrated' functions known from studies of collinear parton distributions
\cite{Zavada:1996kp,Zavada:2001bq,Zavada:2002uz,Efremov:2004tz,Zavada:2006yz,Zavada:2007ww}, 
proven that it satisfies inequalities among TMDs,
and discussed that it yields results consistent with the large-$N_c$ limit,
lattice QCD, and many other models.

In particular, we have also shown that in the approach a relation, which is 
derived from the QCD equations of motion and connects several TMDs and a 
pure twist-3 ('tilde') function, is consequently satisfied in the model.
In our covariant approach with free partons 'consequently' means that the
'tilde'-function is absent. 

We have rederived several known quark model relations among polarized 
leading-twist TMDs \cite{Jakob:1997wg,Avakian:2008dz,Pasquini:2008ax},
and found several new relations so far not observed in models, {\sl without}
assuming SU(6) spin-flavour symmetry. In our approach these relations refer to 
a scale of several ${\rm GeV}^2$.
Whenever previously such relations were observed, the corresponding model 
explicitly made use of the SU(6) symmetry and the results referred to low
hadronic scales \cite{Jakob:1997wg,Avakian:2008dz,Pasquini:2008ax}.
Not all quark models support these relations \cite{Bacchetta:2008af},
and by including gauge-field degrees of freedom \cite{Meissner:2007rx}
such relations are definitely spoiled which one expects to be
the case also in QCD. However, it remains to be seen whether in nature
some of these relations could at least be approximately satisfied.

We have also shown that the Wandzura-Wilczek-type approximation,
which allow to approximate the transverse moments of $g_{1T}^{\perp q}$ and
$h_{1L}^{\perp q}$ in terms of respectively $g_1^q(x)$ and $h_1^q(x)$
are valid in the model upon the neglect of quark mass terms.
In QCD these relations are valid if one in addition neglects also
pure twist-3 terms \cite{Avakian:2007mv}.

As an interesting digression, we have discussed how the covariant
model framework can be used to formulate the non-relativistic limit 
for TMDs, and derived the non-relativistic limit results for all
leading-twist T-even TMDs. In the non-relativistic limit all these 
TMDs are non-zero, however, their transverse moments vanish.
Interestingly the non-relativistic approach is consistent with 
the basic features of the relativistic model calculations.

In this work we focussed on the general aspects of TMDs in the model.
Further consequences for TMDs due to the parton intrinsic motion, 
as well as phenomenological applications 
(see \cite{Efremov:2008mp} for first  results)
will be discussed elsewhere.

%\newpage

%===================  ACKNOWLEDGMENTS ================================
 \vspace{0.5cm}

 \noindent{\bf Acknowledgements.}
 A.~E.\ and O.~T.\ are supported 
 %by the Grants RFBR 06-02-16215 and 07-02-91557, RF MSE RNP.2.2.2.2.6546 (MIREA)
  by the Grants RFBR 09-02-01149 and 07-02-91557, RF MSE RNP.2.2.2.2.6546 (MIREA)
 and by the Heisenberg-Landau
 and (also P.Z.) Votruba-Blokhitsev Programs of JINR.
 P.~Z.\ is supported by the project AV0Z10100502 of the Academy of
 Sciences of the Czech Republic.

\appendix
%===================  APPENDIX: INEQUALITIES =========================
\section{Proof of inequalities}
\label{App:inequalities}

In this Appendix we prove that the inequalities (\ref{Ineq:II}--\ref{Ineq:IV})
are satisfied, if one assumes SU(6) symmetry and if one assumes all TMDs to be
described in terms of the same covariant distribution $J(p^0)$ normalized as 
$\int\di^3 p\,J(p^0)=1$. Then, in SU(6) 
with $N_c=3$ denoting the number of colours \cite{Karl:1984cz},
the TMDs of definite flavour are given by
\ba
    f_1^q(x) = N_q \,f_1(x) \,, && N_u =    \frac{N_c+1}{2} \,,\;\;
                                   N_d =    \frac{N_c-1}{2}  \nonumber\\ 
    g_1^q(x) = P_q \,g_1(x) \,, && P_u =    \frac{N_c+5}{6} \,, \;\;
                                   P_d = -\,\frac{N_c-1}{6} \,,
    \;\;\;\mbox{and analog}\;\;
    g_{1T}^\perp , \, h_1, \, h_{1L}^\perp, \, h_{1T}^\perp.
    \label{Eq-app:assume-SU6}
\ea
The 'flavour-less' functions introduced in (\ref{Eq-app:assume-SU6}) are given 
respectively by Eq.~(\ref{Eq:comp-f1}) with $G(p^0)$ replaced by $J(p^0)$, and
by Eqs.~(\ref{Eq:comp-g1}--\ref{Eq:comp-h1Tperp}) with $H(p^0)$ replaced by 
$J(p^0)$. We immediately see that $g_1(x,p_T) \le f_1(x,p_T)$ and
$h_1(x,p_T) \le f_1(x,p_T)$. Since $|P_q| < N_q$, 
this means that the 'trivial' inequalities 
$|g_1^q(x,p_T)| \le f_1^q(x,p_T)$ and $|h_1^q(x,p_T)| \le f_1^q(x,p_T)$ hold. 

Using the notation of the 'unintegrated' transverse moment of a TMD 
introduced in (\ref{Eq:def-transv-mom}) we obtain the following 
{\sl equalities} among the 'bare' (flavourless) functions
\ba
        f_1(x,p_T)+g_1(x,p_T) &=& \;2\,h_1(x,p_T) \label{Eq-app:rel-I}
	\phantom{\frac11}\\
        f_1(x,p_T)-g_1(x,p_T) &=& -2\,h_{1T}^{\perp(1)}(x,p_T) 
        \label{Eq-app:rel-II}\\
        g_{1T}^{\perp(1)}(x,p_T)^2 = h_{1L}^{\perp(1)}(x,p_T)^2 &=&
        \frac{\vec{p}_T^{\:2}}{4M^2}\biggl(f_1(x,p_T)^2-g_1(x,p_T)^2\biggr) 
        \label{Eq-app:rel-III}
\ea

The relations (\ref{Eq-app:rel-I},~\ref{Eq-app:rel-II}) 
among the bare distributions were discussed previously 
in various models \cite{Avakian:2008dz,Pasquini:2008ax}. 
It is important to notice that even if one assumes SU(6) symmetry, 
the relations (\ref{Eq-app:rel-I},~\ref{Eq-app:rel-II})
among bare distributions do not need to imply relations among
TMDs of definite flavour, though it is the case in the bag and 
constituent-quark models \cite{Avakian:2008dz,Pasquini:2008ax}.
But the spectator model of Ref.~\cite{Jakob:1997wg} provides a counter-example:
there the bare TMDs satisfy (\ref{Eq-app:rel-I},~\ref{Eq-app:rel-II}), but
the flavoured TMDs constructed from the do not.

The equalities  (\ref{Eq-app:rel-I}--\ref{Eq-app:rel-III})
do not mean that the inequalities (\ref{Ineq:I}--\ref{Ineq:IV})
are saturated. That would be the case, in SU(6), only for TMDs 
of s-quarks in $\Lambda^0$ where $N_s=P_s=1$, see \cite{Barone:2001sp} 
where the Soffer bound \cite{Soffer:1994ww} was discussed. 
For the nucleon in SU(6) we have $|P_q|< N_q$, and 
the equalities  (\ref{Eq-app:rel-I}--\ref{Eq-app:rel-III}) lead to real
(never saturated) inequalities (\ref{Ineq:I}--\ref{Ineq:IV}).
(We recall that T-odd distributions are absent in our approach.)

Thus, we conlcude that the inequalities are manifestly satisfied 
in our approach --- if one assumes SU(6) symmetry. If one does not, 
the positivity conditions (\ref{Ineq:I}--\ref{Ineq:IV}) 'translate'
into certain constraints among the covariant momentum distributions
$G(p^0)$ and $H(p^0)$, see Ref.~\cite{Efremov:2004tz}.

%===================  APPENDIX: RELATIONS AMONG TMDs IN SU(6) ========
\section{Relations among TMDs in SU(6)}
\label{App:relations-in-SU6}

In this Appendix we discuss  relations among TMDs that
are obtained in our approach under the assumptions of SU(6) symmetry
and that at all TMDs are characterized in terms of the same covariant
momentum distribution $J(p^0)$, see App.~\ref{App:inequalities}.
From Eqs.~(\ref{Eq-app:assume-SU6}--\ref{Eq-app:rel-III}) we obtain 
the following relations among TMDs with definite flavour
\ba
        \frac{P_q}{N_q}\,f^q_1(x,p_T)+g_1^q(x,p_T) &=& 2h_1^q(x,p_T) 
	\label{Eq-app:rel-I-SU6}\\
        \frac{P_q}{N_q}\,f_1^q(x,p_T)-g_1^q(x,p_T) &=&-2h_{1T}^{\perp(1)q}(x,p_T)
        \label{Eq-app:rel-II-SU6}\\
        g_{1T}^{\perp(1)q}(x,p_T)^2 = h_{1L}^{\perp(1)q}(x,p_T)^2 &=&
        \frac{\vec{p}_T^{\:2}}{4M^2}\biggl(\frac{P_q^2}{N_q^2}
	\,f_1^q(x,p_T)^2-g_1^q(x,p_T)^2\biggr) 
        \label{Eq-app:rel-III-SU6}
\ea
We remark that the relations 
(\ref{Eq-app:rel-I-SU6},~\ref{Eq-app:rel-II-SU6}) hold in the bag
\cite{Avakian:2008dz} and constituent-quark \cite{Pasquini:2008ax} 
model, but not in spectator models \cite{Jakob:1997wg,Bacchetta:2008af}.
Integrated versions of (\ref{Eq-app:rel-I-SU6}) were discussed previously
in \cite{Jaffe:1991ra,Barone:2001sp,Pasquini:2005dk}.

The assumption of SU(6) symmetry by itself is a phenomenologically 
well-motivated concept especially in the valence-$x$ region, 
see \cite{Boffi:2009sh} for a recent discussion in the context of TMDs.
However, in our approach this is not yet a sufficient condition for the 
relations (\ref{Eq-app:rel-I-SU6}--\ref{Eq-app:rel-III-SU6}) to be valid.
{\sl In addition} to SU(6) symmetry, we have to assume here that the
covariant momentum distribution $J(p^0)$ appears in all TMDs.

This fully supports the observation \cite{Avakian:2008dz}, that SU(6) symmetry 
in a quark model {\sl alone} is not a sufficient condition for this kind of 
relations to hold. Another SU(6) symmetric model, which in general does not 
support (\ref{Eq-app:rel-I-SU6}--\ref{Eq-app:rel-III-SU6}),
is the spectator model of \cite{Jakob:1997wg} --- though upon an additional 
assumption (large-$N_c$ limit) they hold there, too \cite{Avakian:2008dz}.

%===================  APPENDIX: PROOF OF WW-TYPE-RELATION ============
\section{Proof of the WW-type relation Eq.~(\ref{Eq:WW-approx-h1L})}
\label{App:proof-of-WW-type-relation}

In this Appendix we present two independent proofs of 
Eq.~(\ref{Eq:WW-approx-h1L}). For the first proof, we use 
the notation of Eq.~(22) in \cite{Zavada:2007ww} to write
the model expressions (\ref{Eq:comp-h1},~\ref{Eq:comp-h1Lperp}) as
\begin{equation}
	h_{1L}^{\bot (1)}(x)=-x^{2}V_{-1}(x)+\frac{x^{3}}{2}V_{-2}(x)
	+{\cal O}\biggl(\frac{m}{M}\biggr),\qquad
	h_{1}(x)=\frac{x^{2}}{2}V_{-2}(x) 
	+{\cal O}\biggl(\frac{m}{M}\biggr) \label{a3} \;.
\end{equation}
Then, exploring the identity
$V_{-1}^{\prime }(x)=\frac{x}{2}\,V_{-2}^{\prime }(x)$
derived Eq.~(24) of \cite{Zavada:2007ww}, we obtain
\begin{equation}
	\left( \frac{h_{1L}^{\bot (1)}(x)}{x^{2}}\right) ^{\prime }
	- \frac{h_{1}(x)}{x^{2}}
	= -V_{-1}^{\prime}(x) +\frac{x}{2}V_{-2}^{\prime }(x)
        + {\cal O}\biggl(\frac{m}{M}\biggr)
	= {\cal O}\biggl(\frac{m}{M}\biggr)
	\label{a6}
\end{equation}
which is equivalent to Eq.~(\ref{Eq:WW-approx-h1L}).

For the second proof we show that $h_{1L}^{\perp(1)q}(x)$ and 
$-x^2\!\int_x^1\frac{\di y}{y^2\;}\,h_1^q(y)$ have the same Mellin moments.
Notice that in the model all TMDs $j^q(x)$ are
well-behaving functions without singularities, have no support
outside the region $x\in[0,\,1]$, and have well-defined Mellin moments 
$\int_0^1\di x\,x^N j^q(x)$ $\forall$ $N=0,\,1,\,2,\,\dots$ 
Therefore Eq.~(\ref{Eq:WW-approx-h1L}) is equivalent to 
\be\label{Eq:to-prove-equiv}
	A_N\equiv
	\int_0^1\di x\biggr(x^N\,h_{1L}^{\perp(1)q}(x)+
	\frac{x^{N+1}}{N+3}\;h_1^q(x)\biggr) 
	= {\cal O}\biggl(\frac{m}{M}\biggr)\;.
\ee
Introducing the notation
$[\di p^3]\equiv\frac{\di^3 p}{p^0}\;\frac{H(p^0)}{p^0+m}$
we write the expressions (\ref{Eq:comp-h1},~\ref{Eq:comp-h1Lperp})
for $h_1^q(x)$ and $h_{1L}^{\perp(1)q}(x)$ as
\ba
        h_{1L}^{\perp(1)q}(x) 
	&=& -\,\frac{M^2}{2}\int[\di p^3]\;
	\delta\left(\frac{p^0-p^1}{M}-x\right)\;x^2\; 
	\frac{\:p^0+p^1}{M} + {\cal O}\biggl(\frac{m}{M}\biggr)\;,
	\label{Eq:comp-h1Lperp-II}\\
        h_1^q(x) 
	&=& \phantom{-}\,
	\frac{M^2}{2}\int[\di p^3]\;
	\delta\left(\frac{p^0-p^1}{M}-x\right)\;x^2 
	 + {\cal O}\biggl(\frac{m}{M}\biggr)\;. \label{Eq:comp-h1-II}
\ea
where $p^0=|\vec{p}\,|$ (recall we neglect $m$). We insert the 
expressions (\ref{Eq:comp-h1Lperp-II},~\ref{Eq:comp-h1-II}) into
Eq.~(\ref{Eq:to-prove-equiv}), interchange the order of the integrations 
over $x$ and $p$, introduce spherical coordinates
% in a somewhat non-standard way with the angle $\theta$ counted 
% from the $1$-axis 
such that $p^1=|\vec{p}\,|\;\cos\theta$,
and obtain
\ba\label{Eq:to-prove-equiv-2}
	A_N 
 &=& 	\frac{1}{2M^{N+1}}\int\{\di^3 p\}\;|\vec{p}\,|^{N+2}\,
	\biggr(-\,(1-\cos\theta)^{N+2}\;(1+\cos\theta)+
	\frac{(1-\cos\theta)^{N+3}}{N+3} \biggr)  
	+ {\cal O}\biggl(\frac{m}{M}\biggr) \,.
\ea
Now, our proof is completed because in (\ref{Eq:to-prove-equiv-2})
the integral over $z\equiv\cos\theta$ is
\be\label{Eq:to-prove-equiv-4}
	\int_{-1}^1\di z\biggr(-\,(1-z)^{N+2}\;(1+z)
	+ \frac{(1-z)^{N+3}}{N+3} \biggr)  
	= \int_{-1}^1\di z\;\frac{\di\;}{\di z}
	\Biggr(\frac{(1-z)^{N+3}(1+z)}{N+3}\Biggr)  
	= 0\;.
\ee

\end{document}